\documentclass[sigconf,natbib=true]{acmart}

\AtBeginDocument{%
  \providecommand\BibTeX{{%
    \normalfont B\kern-0.5em{\scshape i\kern-0.25em b}\kern-0.8em\TeX}}}

\usepackage{acronym}
\usepackage{amsmath} %

\usepackage{amssymb} %
\usepackage{acronym}
\usepackage{enumerate}
\usepackage{amsthm}

\usepackage[inline]{enumitem}

\acrodef{CF}{collaborative filtering}

\setcopyright{acmcopyright}

\acrodef{LTR}{learning to rank}
\acrodef{NDCG}{normalized discounted cumulative gain}
\acrodef{ULTR}{unbiased learning to rank}
\acrodef{IPS}{inverse propensity scoring}

\copyrightyear{2023}
\acmYear{2023}
\setcopyright{rightsretained}
\acmConference[SIGIR '23]{Proceedings of the 46th International ACM SIGIR
Conference on Research and Development in Information Retrieval}{July 23--27,
2023}{Taipei, Taiwan}
\acmBooktitle{Proceedings of the 46th International ACM SIGIR Conference on
Research and Development in Information Retrieval (SIGIR '23), July 23--27,
2023, Taipei, Taiwan}\acmDOI{10.1145/3539618.3594247}
\acmISBN{978-1-4503-9408-6/23/07}

\author{Shashank Gupta}
\orcid{0000-0003-1291-7951}
\affiliation{%
	\institution{University of Amsterdam}
	\city{Amsterdam}
	\country{The Netherlands}
}
\email{s.gupta2@uva.nl}

\author{Philipp Hager}
\orcid{0000-0001-5696-9732}
\affiliation{%
	\institution{University of Amsterdam}
	\city{Amsterdam}
	\country{The Netherlands}
}
\email{p.k.hager@uva.nl}

\author{Jin Huang}
\orcid{0000-0001-9273-9037}
\affiliation{%
	\institution{University of Amsterdam}
	\city{Amsterdam}
	\country{The Netherlands}
}
\email{j.huang2@uva.nl}

\author{Ali Vardasbi}
\orcid{0000-0002-9342-5272}
\affiliation{%
	\institution{University of Amsterdam}
	\city{Amsterdam}
	\country{The Netherlands}
}
\email{a.vardasbi@uva.nl}

\author{Harrie Oosterhuis}
\orcid{0000-0002-0458-9233}
\affiliation{%
	\institution{Radboud University}
	\city{Nijmegen}
	\country{The Netherlands}
}
\email{harrie.oosterhuis@ru.nl}

\begin{document}

\title[Recent Advances in the Foundations and Applications of Unbiased Learning to Rank]{Recent Advances in the Foundations and Applications\\ of Unbiased Learning to Rank}

\renewcommand{\shortauthors}{Gupta, et al.}

\begin{abstract}
Since its inception, the field of \ac{ULTR} has remained very active and has seen several impactful advancements in recent years.
This tutorial provides both an introduction to the core concepts of the field and an overview of recent advancements in its foundations along with several applications of its methods.

The tutorial is divided into four parts:
Firstly, we give an overview of the different forms of bias that can be addressed with \ac{ULTR} methods.
Secondly, we present a comprehensive discussion of the latest estimation techniques in the \ac{ULTR} field.
Thirdly, we survey published results of \ac{ULTR} in real-world applications.
Fourthly, we discuss the connection between \ac{ULTR} and fairness in ranking.
We end by briefly reflecting on the future of \ac{ULTR} research and its applications.

This tutorial is intended to benefit both researchers and industry practitioners who are interested in developing new \ac{ULTR} solutions or utilizing them in real-world applications.
\end{abstract}



\settopmatter{printacmref=true}

\settopmatter{printfolios=true}
\maketitle

\section{Motivation}
\Ac{LTR} algorithms are the cornerstone of modern search and recommender systems. Traditionally, \acs{LTR} algorithms were based on supervised learning from manually-graded relevance judgments. However, obtaining relevance judgments is costly and often not aligned with actual user preferences~\cite{chapelle2011yahoo,sanderson2010user}. In contrast, click data is cheaper to collect and is generally better aligned with user intents~\citep{joachims2002optimizing}. However, click data is usually a heavily biased signal of user preference~\citep{craswell2008experimental,joachims2017unbiased,agarwal2019addressing} which the field of \acf{ULTR} aims to mitigate~\cite{oosterhuis2020learning}.

Previous tutorials focused on introducing the fundamentals of the field to researchers and practitioners in the information retrieval and recommender system communities~\cite{oosterhuis2020unbiased,ai2018unbiased,lucchese2019learning}.
While very relevant at the time, the field of \ac{ULTR} has matured significantly, and fundamental advancements have been made since then.
At the time of the last tutorials, the primarily studied interaction bias was position bias~\cite{craswell2008experimental,joachims2017unbiased,wang2018position}.
Since then, the community has addressed new interaction biases, including trust bias~\cite{agarwal2019addressing,vardasbi2020inverse}, item selection bias~\cite{oosterhuis2020policy}, contextual bias~\cite{wu2021unbiased,zhuang2021cross,zheng2022cbr,yan2022revisiting}, and cascading position bias~\cite{kiyohara2022doubly,vardasbi2020cascade}.
For correcting biases, the method most commonly used was \ac{IPS}.
However, it is now known that \ac{IPS} is not effective in correcting for all forms of interaction biases~\cite{vardasbi2020inverse,oosterhuis2022reaching}.
Hence, several new and fundamental estimation techniques have been developed to overcome the limitations of \ac{IPS}, for instance, affine-corrections~\cite{vardasbi2020inverse,oosterhuis2021unifying} and doubly-robust estimation~\cite{oosterhuis2023doubly}, which can both be seen as extensions of \ac{IPS} for \ac{ULTR}.
Furthermore, estimation methods that are fundamentally different from \ac{IPS} have also been proposed, such as two-tower models~\cite{yan2022revisiting, zhuang2021cross, guo2019pal} and causal-inference-based methods~\cite{ovaisi2021propensity, zhao2022unbiased, tian2020hte}.
While many \ac{ULTR} methods focus on mitigating bias in historic datasets, the area of online learning to rank mitigates biases while directly interacting with users~\cite{yue2009interactively,schuth2016multileave,oosterhuis2018differentiable}.
A recent line of work addresses both online and offline settings with methods that can be applied to either setting and thereby, aims to unify the \ac{ULTR} field~\cite{oosterhuis2020taking,oosterhuis2021unifying}.

Recently, \ac{LTR} has also seen significant growth from the application side~\cite{agarwal2019addressing,zhuang2021cross,wu2021unbiased,hu2019unbiased,block2022counterfactual}, including fair \ac{LTR}~\cite{singh2019policy, singh2021fairness,yadav2021policy}. The focus of the previous tutorials was on the fundamentals of \ac{ULTR} with a limited emphasis on practical applications. While the focus of \ac{LTR} was traditionally on relevance ranking, it is now commonly acknowledged that optimizing for relevance alone can result in unfairness issues~\cite{biega2018equity, singh2018fairness, yang2021maximizing}.
In this regard, we believe that the objective of a similar area, such as fair \ac{LTR}, aligns with \ac{ULTR}'s mission, which is to provide fair and unbiased rankings to the user.
To scale up to large-scale applications, fair \ac{LTR} work relies on unbiased \ac{LTR}~\citep{yadav2021policy, sarvi2021understanding, vardasbi2022probabilistic}, and we hope that our tutorial will encourage further exploration in this area. 

Given these significant advancements in the area of \ac{ULTR} and the increased applications of its methodology, we believe it is the right time to provide an overview of the state-of-the-art of the field.
Hereby, we aim to benefit both academic researchers and industry practitioners who are either interested in developing new \ac{ULTR} solutions or utilizing them in their applications.

\section{Objectives}
The tutorial is based on the following two main objectives:
\begin{itemize}[leftmargin=*]
    \item To motivate and introduce the fundamental concepts of ULTR to academics or practitioners who are new to the topic.
    \item To provide a comprehensive overview of the important recent developments to the foundations and applications of ULTR, that are useful to both newcomers and experts in the field.
\end{itemize}
Furthermore, we aim for the following additional objectives:
 \begin{itemize}[leftmargin=*]
 \item Provide the most up-to-date explanation of the mathematical foundations of the \ac{ULTR} field, covering the different forms of bias that can and cannot be corrected for and the latest estimation techniques.
 Thereby, for researchers in the ULTR field, this should provide them with a strong basis to perform future research.
\item Present an in-depth survey of real-world applications of \ac{ULTR} so that practitioners can have a realistic expectation of the potential impact of applying \ac{ULTR}. 
\item Motivate and stimulate cross-disciplinary research, by enabling researchers from other areas to understand how \ac{ULTR} could be useful for them.
In particular, we will highlight the connection with the topic of \emph{fairness in ranking}.
\end{itemize}

\section{Relevance to the Information Retrieval Community}
The area of \ac{ULTR} has grown significantly in the last couple of years, with several fundamental contributions and diverse IR applications.

The earliest tutorial in the area was by \citet{joachims2016counterfactual}, where they introduced counterfactual learning in the context of search and recommendation. Recent tutorials on counterfactual evaluation and learning have focused mostly on bandit feedback data~\cite{saito2022counterfactual,saito2021counterfactual}. In the context of recommender systems,~\citet{chen2021bias} introduce biases and debiasing strategies.  

For \ac{LTR} specifically, there have been tutorials introducing offline \ac{ULTR}~\cite{ai2018unbiased,oosterhuis2020unbiased,lucchese2019learning} and online \ac{LTR}~\cite{grotov2016online}.
However, to the best of our knowledge, no existing tutorial covers the important advancements in the \ac{ULTR} field that has been made in the last three years, nor their recent applications, including fair \ac{LTR}~\citep{singh2019policy, yadav2021policy}.

\section{Format and Detailed Schedule}
The tutorial will consist of three hours, excluding breaks, with the following schedule:

\noindent \textbf{Preliminaries (20 minutes)} The first session focuses on the preliminaries; we discuss the basics of supervised \ac{LTR} and some of the earliest works in position-bias and counterfactual \ac{LTR}.
\begin{itemize}[leftmargin=*]
    \item \textbf{Learning to rank basics (5 minutes)}: Discuss basics of supervised \ac{LTR} by introducing pointwise, pairwise, and listwise \ac{LTR} methods and the concept of learning from user interactions. 
    \item \textbf{Position bias (5 minutes)}: Discuss position bias that arises when applying traditional \ac{LTR} methods on user clicks~\cite{craswell2008experimental}.
    \item \textbf{Counterfactual \ac{LTR} (10 minutes)}: Introduce the basics of counterfactual \ac{LTR} for learning from user feedback data with position bias~\cite{joachims2017unbiased}.
\end{itemize}

\noindent \textbf{Biases (40 minutes)} In this session, we cover the recent advances in the types of interaction bias that can be tackled with \ac{ULTR} methods beyond position bias.
\begin{itemize}[leftmargin=*]
        \item \textbf{Trust bias (10 minutes)}: We discuss trust bias, where users are likely to click on items in top positions regardless of item relevance because they trust the search engine~\cite{agarwal2019addressing,vardasbi2020inverse}. 
        \item \textbf{Item Selection Bias (10 minutes)}: We discuss item selection bias, where users can only interact with a fixed set of $k$ items, and items outside the top-k position have zero chances of exposure~\cite{oosterhuis2020policy}. 
        \item \textbf{Contextual Bias (10 minutes)}: We discuss contextual bias, where the item's click probability is affected by its surrounding items on the display page~\cite{wu2021unbiased,zhuang2021cross,zheng2022cbr,yan2022revisiting}.
        \item \textbf{Cascading Position Bias (10 minutes)}: Under the cascade model~\cite{craswell2008experimental}, the position bias of an item depends not only on the rank an item is displayed at (as many works in \ac{ULTR} assume~\cite{craswell2008experimental,joachims2017unbiased,wang2018position}), but also on the relevance of the items the user has inspected before~\cite{kiyohara2022doubly,vardasbi2020cascade}. Thus, cascading position bias is often a more realistic assumption of user behavior, e.g., when users tend to stop searching after finding the first relevant result~\cite{craswell2008experimental}.
\end{itemize}
\textbf{Novel Estimation Methods (70 minutes)} The most prevalent estimation technique for correcting bias from user interaction data is \ac{IPS}, introduced in the seminal works by \citet{wang2016learning} and \citet{joachims2017unbiased}.  However, recently there have been fundamental contributions in the \ac{ULTR} field with respect to novel estimation techniques. In this session, we will discuss the recent estimation techniques in detail, as per the following schedule:
\vspace{-0.75mm}
\begin{itemize}[leftmargin=*]
        \item \textbf{Affine Correction Method (10 minutes)}: We discuss the affine correction method introduced by \citet{vardasbi2020inverse} as an extension to \ac{IPS}, which they prove to be ineffective when correcting for trust-bias.  
        \item \textbf{Doubly Robust Estimation (10 minutes)}: Despite its popularity in the offline bandit learning literature~\citep{dudik2011doubly,jiang2016doubly,huang2020importance,saito2020doubly,wang2019doubly}, doubly robust estimation methods for position bias correction in \ac{LTR} were only proposed recently~\cite{oosterhuis2023doubly}. We discuss this fundamental contribution to the area which overcomes some of the theoretical and practical disadvantages of \ac{IPS}~\cite{oosterhuis2023doubly}. 
        \item \textbf{Online \& Counterfactual methods (10 minutes)}: Online learning methods are an alternative class of methods to counter biases in \ac{LTR}~\cite{yue2009interactively,schuth2016multileave,oosterhuis2018differentiable}, where the user preferences are learned in an online/interactive fashion, as opposed to purely from offline data. Recently, \citet{oosterhuis2021unifying} argued that the field of \ac{ULTR} can benefit from using online learning to rank via a novel online intervention-aware counterfactual estimator. Online learning has also been used to collect additional data from the logging policy to minimize the variance of the counterfactual estimate of a new ranking policy~\cite{oosterhuis2020taking}.  
        \item \textbf{Safe Counterfactual Policy Learning (10 minutes)}: \Ac{ULTR} relies on exposure-based \ac{IPS}, which can provide unbiased and consistent estimates but often suffers from high variance.
        Especially when little click data is available, this variance can cause \ac{ULTR} to learn sub-optimal ranking behavior, which can subsequently bring significant risks of a negative user experience. Recently, \citet{gupta-2023-safe} introduced a risk-aware \ac{ULTR} method with a novel exposure-based concept of risk regularization with strong theoretical guarantees for safe deployment. Thereby, it averts the risk of a learned policy being worse than the production ranking system. 
        \item \textbf{Marginalized propensities (10 minutes)}:
        \ac{IPS}-based estimators rely heavily on the full-support assumption for being unbiased, i.e., the logging policy is expected to have non-zero probability over all actions assumed under the new policy~\cite{li2011unbiased}. %
        This assumption can be restrictive in a setting where the action space is large, as is typical in many real-world applications. To deal with large-action space, \citet{saito2022off} introduced the marginal importance sampling method, which leverages so-called action embeddings to generalize to a large action space.
        \item \textbf{Two-tower Models (10 minutes):}
        So far, the focus of \ac{ULTR} was primarily on identifying and developing novel bias correction methods, with limited focus on model design. 
        Recently, with the introduction of two-tower networks, this trend has been slowly changing~\cite{yan2022revisiting,zhuang2021cross,guo2019pal}.
        We discuss the two-tower family of correction methods. 
        \item \textbf{Non-propensity based methods (10 minutes):}
        The primary estimation techniques for bias correction in \ac{ULTR} are based on \ac{IPS}~\cite{joachims2017unbiased,wang2016learning,agarwal2019addressing,vardasbi2020inverse,oosterhuis2022doubly}.
        Despite its advantages, \ac{IPS} often suffers from high-variance~\cite{oosterhuis2023doubly}.
        Recently, the field of \ac{ULTR} has progressed from relying primarily on \ac{IPS} to other non-IPS-based bias correction techniques, such as causal-inference-based methods like Heckerman's method~\cite{ovaisi2020correcting} and the causal likelihood decomposition method~\cite{zhao2022unbiased}.
\end{itemize}

\noindent \textbf{Survey Applications (20 minutes)}
A significant number of works apply \ac{ULTR} methods in real-world settings.
We discuss the different settings explored by existing work, the practical changes they require, and the reported performance impact.
A brief overview of the settings that will be covered:
\begin{itemize}[leftmargin=*]
    \item  \textbf{Top-K ranking}: Application of \ac{ULTR} in top-k rankings~\cite{agarwal2019addressing,zhuang2021cross}. 
    \item \textbf{Feeds recommendation}: \ac{ULTR} for product feeds recommendation, with user interactions logged from a 2D grid-based user interface~\cite{wu2021unbiased,zhuang2021cross}.
    \item \textbf{Job recommendation}: Job recommendations using \ac{ULTR}~\cite{chen2019correcting}.
    \item \textbf{Fair \ac{LTR}}: Fair policy learning for article recommendation~\cite{singh2021fairness}.
\end{itemize}

\noindent \textbf{From Unbiased to Fair \ac{LTR} (20 minutes)}
Traditionally, fair \ac{LTR} has relied on manual relevance judgments for learning fair ranking policies~\citep{biega2018equity,singh2018fairness}.
Similar to the arguments in favor of click-based learning for relevance rankings, fair \ac{LTR} needs to adopt click data for its widespread application.
In this part of the tutorial, we discuss applications of \ac{ULTR} for fair policy learning~\citep{yadav2021policy,morik2020controlling}.

\noindent \textbf{Conclusion and Future Work (10 minutes)}
We close by summarizing the main points discussed in the tutorial, in addition, we also discuss some important limitations of the existing overarching counterfactual approach in the \ac{ULTR} field~\cite{oosterhuis2022reaching} and some promising avenues for future research. 

\section{Supplied Material}
The tutorial slides along with an annotated bibliographic compilation of references, open reference code from related work, and a public video recording of the presentation are available at the \href{https://sites.google.com/view/sigir-2023-tutorial-ultr}{\color{blue}tutorial website}.

\balance
\bibliographystyle{ACM-Reference-Format}
\bibliography{bibliography}


\begin{thebibliography}{58}


\ifx \showCODEN    \undefined \def \showCODEN     #1{\unskip}     \fi
\ifx \showDOI      \undefined \def \showDOI       #1{#1}\fi
\ifx \showISBNx    \undefined \def \showISBNx     #1{\unskip}     \fi
\ifx \showISBNxiii \undefined \def \showISBNxiii  #1{\unskip}     \fi
\ifx \showISSN     \undefined \def \showISSN      #1{\unskip}     \fi
\ifx \showLCCN     \undefined \def \showLCCN      #1{\unskip}     \fi
\ifx \shownote     \undefined \def \shownote      #1{#1}          \fi
\ifx \showarticletitle \undefined \def \showarticletitle #1{#1}   \fi
\ifx \showURL      \undefined \def \showURL       {\relax}        \fi
\providecommand\bibfield[2]{#2}
\providecommand\bibinfo[2]{#2}
\providecommand\natexlab[1]{#1}
\providecommand\showeprint[2][]{arXiv:#2}

\bibitem[\protect\citeauthoryear{Agarwal, Wang, Li, Bendersky, and
  Najork}{Agarwal et~al\mbox{.}}{2019}]%
        {agarwal2019addressing}
\bibfield{author}{\bibinfo{person}{Aman Agarwal}, \bibinfo{person}{Xuanhui
  Wang}, \bibinfo{person}{Cheng Li}, \bibinfo{person}{Michael Bendersky}, {and}
  \bibinfo{person}{Marc Najork}.} \bibinfo{year}{2019}\natexlab{}.
\newblock \showarticletitle{Addressing Trust Bias for Unbiased
  Learning-to-rank}. In \bibinfo{booktitle}{\emph{The World Wide Web
  Conference}}. \bibinfo{pages}{4--14}.
\newblock


\bibitem[\protect\citeauthoryear{Ai, Mao, Liu, and Croft}{Ai
  et~al\mbox{.}}{2018}]%
        {ai2018unbiased}
\bibfield{author}{\bibinfo{person}{Qingyao Ai}, \bibinfo{person}{Jiaxin Mao},
  \bibinfo{person}{Yiqun Liu}, {and} \bibinfo{person}{W~Bruce Croft}.}
  \bibinfo{year}{2018}\natexlab{}.
\newblock \showarticletitle{Unbiased learning to rank: Theory and practice}. In
  \bibinfo{booktitle}{\emph{Proceedings of the 27th ACM International
  Conference on Information and Knowledge Management}}.
  \bibinfo{pages}{2305--2306}.
\newblock


\bibitem[\protect\citeauthoryear{Biega, Gummadi, and Weikum}{Biega
  et~al\mbox{.}}{2018}]%
        {biega2018equity}
\bibfield{author}{\bibinfo{person}{Asia~J Biega}, \bibinfo{person}{Krishna~P
  Gummadi}, {and} \bibinfo{person}{Gerhard Weikum}.}
  \bibinfo{year}{2018}\natexlab{}.
\newblock \showarticletitle{Equity of attention: Amortizing individual fairness
  in rankings}. In \bibinfo{booktitle}{\emph{The 41st international acm sigir
  conference on research \& development in information retrieval}}.
  \bibinfo{pages}{405--414}.
\newblock


\bibitem[\protect\citeauthoryear{Block, Kidambi, Hill, Joachims, and
  Dhillon}{Block et~al\mbox{.}}{2022}]%
        {block2022counterfactual}
\bibfield{author}{\bibinfo{person}{Adam Block}, \bibinfo{person}{Rahul
  Kidambi}, \bibinfo{person}{Daniel~N Hill}, \bibinfo{person}{Thorsten
  Joachims}, {and} \bibinfo{person}{Inderjit~S Dhillon}.}
  \bibinfo{year}{2022}\natexlab{}.
\newblock \showarticletitle{Counterfactual Learning To Rank for
  Utility-Maximizing Query Autocompletion}.
\newblock \bibinfo{journal}{\emph{arXiv preprint arXiv:2204.10936}}
  (\bibinfo{year}{2022}).
\newblock


\bibitem[\protect\citeauthoryear{Chapelle and Chang}{Chapelle and
  Chang}{2011}]%
        {chapelle2011yahoo}
\bibfield{author}{\bibinfo{person}{Olivier Chapelle} {and} \bibinfo{person}{Yi
  Chang}.} \bibinfo{year}{2011}\natexlab{}.
\newblock \showarticletitle{Yahoo! Learning to Rank Challenge Overview}. In
  \bibinfo{booktitle}{\emph{Proceedings of the learning to rank challenge}}.
  PMLR, \bibinfo{pages}{1--24}.
\newblock


\bibitem[\protect\citeauthoryear{Chen, Wang, Feng, and He}{Chen
  et~al\mbox{.}}{2021}]%
        {chen2021bias}
\bibfield{author}{\bibinfo{person}{Jiawei Chen}, \bibinfo{person}{Xiang Wang},
  \bibinfo{person}{Fuli Feng}, {and} \bibinfo{person}{Xiangnan He}.}
  \bibinfo{year}{2021}\natexlab{}.
\newblock \showarticletitle{Bias Issues and Solutions in Recommender System:
  Tutorial on the RecSys 2021}. In \bibinfo{booktitle}{\emph{Proceedings of the
  15th ACM Conference on Recommender Systems}}. \bibinfo{pages}{825--827}.
\newblock


\bibitem[\protect\citeauthoryear{Chen, Ai, Jayasinghe, and Croft}{Chen
  et~al\mbox{.}}{2019}]%
        {chen2019correcting}
\bibfield{author}{\bibinfo{person}{Ruey-Cheng Chen}, \bibinfo{person}{Qingyao
  Ai}, \bibinfo{person}{Gaya Jayasinghe}, {and} \bibinfo{person}{W~Bruce
  Croft}.} \bibinfo{year}{2019}\natexlab{}.
\newblock \showarticletitle{Correcting for recency bias in job recommendation}.
  In \bibinfo{booktitle}{\emph{Proceedings of the 28th ACM international
  conference on information and knowledge management}}.
  \bibinfo{pages}{2185--2188}.
\newblock


\bibitem[\protect\citeauthoryear{Craswell, Zoeter, Taylor, and Ramsey}{Craswell
  et~al\mbox{.}}{2008}]%
        {craswell2008experimental}
\bibfield{author}{\bibinfo{person}{Nick Craswell}, \bibinfo{person}{Onno
  Zoeter}, \bibinfo{person}{Michael Taylor}, {and} \bibinfo{person}{Bill
  Ramsey}.} \bibinfo{year}{2008}\natexlab{}.
\newblock \showarticletitle{An Experimental Comparison of Click Position-bias
  Models}. In \bibinfo{booktitle}{\emph{Proceedings of the 2008 international
  conference on web search and data mining}}. \bibinfo{pages}{87--94}.
\newblock


\bibitem[\protect\citeauthoryear{Dud{\'\i}k, Langford, and Li}{Dud{\'\i}k
  et~al\mbox{.}}{2011}]%
        {dudik2011doubly}
\bibfield{author}{\bibinfo{person}{Miroslav Dud{\'\i}k}, \bibinfo{person}{John
  Langford}, {and} \bibinfo{person}{Lihong Li}.}
  \bibinfo{year}{2011}\natexlab{}.
\newblock \showarticletitle{Doubly robust policy evaluation and learning}.
\newblock \bibinfo{journal}{\emph{arXiv preprint arXiv:1103.4601}}
  (\bibinfo{year}{2011}).
\newblock


\bibitem[\protect\citeauthoryear{Grotov and De~Rijke}{Grotov and
  De~Rijke}{2016}]%
        {grotov2016online}
\bibfield{author}{\bibinfo{person}{Artem Grotov} {and} \bibinfo{person}{Maarten
  De~Rijke}.} \bibinfo{year}{2016}\natexlab{}.
\newblock \showarticletitle{Online learning to rank for information retrieval:
  Sigir 2016 tutorial}. In \bibinfo{booktitle}{\emph{Proceedings of the 39th
  International ACM SIGIR conference on Research and Development in Information
  Retrieval}}. \bibinfo{pages}{1215--1218}.
\newblock


\bibitem[\protect\citeauthoryear{Guo, Yu, Liu, Tang, and Zhang}{Guo
  et~al\mbox{.}}{2019}]%
        {guo2019pal}
\bibfield{author}{\bibinfo{person}{Huifeng Guo}, \bibinfo{person}{Jinkai Yu},
  \bibinfo{person}{Qing Liu}, \bibinfo{person}{Ruiming Tang}, {and}
  \bibinfo{person}{Yuzhou Zhang}.} \bibinfo{year}{2019}\natexlab{}.
\newblock \showarticletitle{PAL: a position-bias aware learning framework for
  CTR prediction in live recommender systems}. In
  \bibinfo{booktitle}{\emph{Proceedings of the 13th ACM Conference on
  Recommender Systems}}. \bibinfo{pages}{452--456}.
\newblock


\bibitem[\protect\citeauthoryear{Gupta, Oosterhuis, and de~Rijke}{Gupta
  et~al\mbox{.}}{2023}]%
        {gupta-2023-safe}
\bibfield{author}{\bibinfo{person}{Shashank Gupta}, \bibinfo{person}{Harrie
  Oosterhuis}, {and} \bibinfo{person}{Maarten de Rijke}.}
  \bibinfo{year}{2023}\natexlab{}.
\newblock \showarticletitle{Safe Deployment for Counterfactual Learning to Rank
  with Exposure-Based Risk Minimization}. In \bibinfo{booktitle}{\emph{SIGIR
  2023: 46th international ACM SIGIR Conference on Research and Development in
  Information Retrieval}}. \bibinfo{publisher}{ACM}.
\newblock


\bibitem[\protect\citeauthoryear{Hu, Wang, Peng, and Li}{Hu
  et~al\mbox{.}}{2019}]%
        {hu2019unbiased}
\bibfield{author}{\bibinfo{person}{Ziniu Hu}, \bibinfo{person}{Yang Wang},
  \bibinfo{person}{Qu Peng}, {and} \bibinfo{person}{Hang Li}.}
  \bibinfo{year}{2019}\natexlab{}.
\newblock \showarticletitle{Unbiased lambdamart: an unbiased pairwise
  learning-to-rank algorithm}. In \bibinfo{booktitle}{\emph{The World Wide Web
  Conference}}. \bibinfo{pages}{2830--2836}.
\newblock


\bibitem[\protect\citeauthoryear{Huang and Jiang}{Huang and Jiang}{2020}]%
        {huang2020importance}
\bibfield{author}{\bibinfo{person}{Jiawei Huang} {and} \bibinfo{person}{Nan
  Jiang}.} \bibinfo{year}{2020}\natexlab{}.
\newblock \showarticletitle{From importance sampling to doubly robust policy
  gradient}. In \bibinfo{booktitle}{\emph{International Conference on Machine
  Learning}}. PMLR, \bibinfo{pages}{4434--4443}.
\newblock


\bibitem[\protect\citeauthoryear{Jiang and Li}{Jiang and Li}{2016}]%
        {jiang2016doubly}
\bibfield{author}{\bibinfo{person}{Nan Jiang} {and} \bibinfo{person}{Lihong
  Li}.} \bibinfo{year}{2016}\natexlab{}.
\newblock \showarticletitle{Doubly robust off-policy value evaluation for
  reinforcement learning}. In \bibinfo{booktitle}{\emph{International
  Conference on Machine Learning}}. PMLR, \bibinfo{pages}{652--661}.
\newblock


\bibitem[\protect\citeauthoryear{Joachims}{Joachims}{2002}]%
        {joachims2002optimizing}
\bibfield{author}{\bibinfo{person}{Thorsten Joachims}.}
  \bibinfo{year}{2002}\natexlab{}.
\newblock \showarticletitle{Optimizing Search Engines Using Clickthrough Data}.
  In \bibinfo{booktitle}{\emph{Proceedings of the eighth ACM SIGKDD
  international conference on Knowledge discovery and data mining}}.
  \bibinfo{pages}{133--142}.
\newblock


\bibitem[\protect\citeauthoryear{Joachims and Swaminathan}{Joachims and
  Swaminathan}{2016}]%
        {joachims2016counterfactual}
\bibfield{author}{\bibinfo{person}{Thorsten Joachims} {and}
  \bibinfo{person}{Adith Swaminathan}.} \bibinfo{year}{2016}\natexlab{}.
\newblock \showarticletitle{Counterfactual Evaluation and Learning for Search,
  Recommendation and Ad Placement}. In \bibinfo{booktitle}{\emph{Proceedings of
  the 39th International ACM SIGIR conference on Research and Development in
  Information Retrieval}}. \bibinfo{pages}{1199--1201}.
\newblock


\bibitem[\protect\citeauthoryear{Joachims, Swaminathan, and Schnabel}{Joachims
  et~al\mbox{.}}{2017}]%
        {joachims2017unbiased}
\bibfield{author}{\bibinfo{person}{Thorsten Joachims}, \bibinfo{person}{Adith
  Swaminathan}, {and} \bibinfo{person}{Tobias Schnabel}.}
  \bibinfo{year}{2017}\natexlab{}.
\newblock \showarticletitle{Unbiased Learning-to-rank with Biased Feedback}. In
  \bibinfo{booktitle}{\emph{Proceedings of the Tenth ACM International
  Conference on Web Search and Data Mining}}. \bibinfo{pages}{781--789}.
\newblock


\bibitem[\protect\citeauthoryear{Kiyohara, Saito, Matsuhiro, Narita, Shimizu,
  and Yamamoto}{Kiyohara et~al\mbox{.}}{2022}]%
        {kiyohara2022doubly}
\bibfield{author}{\bibinfo{person}{Haruka Kiyohara}, \bibinfo{person}{Yuta
  Saito}, \bibinfo{person}{Tatsuya Matsuhiro}, \bibinfo{person}{Yusuke Narita},
  \bibinfo{person}{Nobuyuki Shimizu}, {and} \bibinfo{person}{Yasuo Yamamoto}.}
  \bibinfo{year}{2022}\natexlab{}.
\newblock \showarticletitle{Doubly robust off-policy evaluation for ranking
  policies under the cascade behavior model}. In
  \bibinfo{booktitle}{\emph{Proceedings of the Fifteenth ACM International
  Conference on Web Search and Data Mining}}. \bibinfo{pages}{487--497}.
\newblock


\bibitem[\protect\citeauthoryear{Li, Chu, Langford, and Wang}{Li
  et~al\mbox{.}}{2011}]%
        {li2011unbiased}
\bibfield{author}{\bibinfo{person}{Lihong Li}, \bibinfo{person}{Wei Chu},
  \bibinfo{person}{John Langford}, {and} \bibinfo{person}{Xuanhui Wang}.}
  \bibinfo{year}{2011}\natexlab{}.
\newblock \showarticletitle{Unbiased Offline Evaluation of
  Contextual-Bandit-based News Article Recommendation Algorithms}. In
  \bibinfo{booktitle}{\emph{Proceedings of the fourth ACM international
  conference on Web Search and Data Mining}}. \bibinfo{pages}{297--306}.
\newblock


\bibitem[\protect\citeauthoryear{Lucchese, Nardini, Pasumarthi, Bruch,
  Bendersky, Wang, Oosterhuis, Jagerman, and de~Rijke}{Lucchese
  et~al\mbox{.}}{2019}]%
        {lucchese2019learning}
\bibfield{author}{\bibinfo{person}{Claudio Lucchese},
  \bibinfo{person}{Franco~Maria Nardini}, \bibinfo{person}{Rama~Kumar
  Pasumarthi}, \bibinfo{person}{Sebastian Bruch}, \bibinfo{person}{Michael
  Bendersky}, \bibinfo{person}{Xuanhui Wang}, \bibinfo{person}{Harrie
  Oosterhuis}, \bibinfo{person}{Rolf Jagerman}, {and} \bibinfo{person}{Maarten
  de Rijke}.} \bibinfo{year}{2019}\natexlab{}.
\newblock \showarticletitle{Learning to rank in theory and practice: from
  gradient boosting to neural networks and unbiased learning}. In
  \bibinfo{booktitle}{\emph{Proceedings of the 42nd International ACM SIGIR
  Conference on Research and Development in Information Retrieval}}.
  \bibinfo{pages}{1419--1420}.
\newblock


\bibitem[\protect\citeauthoryear{Morik, Singh, Hong, and Joachims}{Morik
  et~al\mbox{.}}{2020}]%
        {morik2020controlling}
\bibfield{author}{\bibinfo{person}{Marco Morik}, \bibinfo{person}{Ashudeep
  Singh}, \bibinfo{person}{Jessica Hong}, {and} \bibinfo{person}{Thorsten
  Joachims}.} \bibinfo{year}{2020}\natexlab{}.
\newblock \showarticletitle{Controlling fairness and bias in dynamic
  learning-to-rank}. In \bibinfo{booktitle}{\emph{Proceedings of the 43rd
  international ACM SIGIR conference on research and development in information
  retrieval}}. \bibinfo{pages}{429--438}.
\newblock


\bibitem[\protect\citeauthoryear{Oosterhuis}{Oosterhuis}{2020}]%
        {oosterhuis2020learning}
\bibfield{author}{\bibinfo{person}{Harrie Oosterhuis}.}
  \bibinfo{year}{2020}\natexlab{}.
\newblock \emph{\bibinfo{title}{Learning from User Interactions with Rankings:
  A Unification of the Field}}.
\newblock \bibinfo{thesistype}{Ph.\,D. Dissertation}.
  \bibinfo{school}{Informatics Institute, University of Amsterdam}.
\newblock


\bibitem[\protect\citeauthoryear{Oosterhuis}{Oosterhuis}{2022a}]%
        {oosterhuis2022doubly}
\bibfield{author}{\bibinfo{person}{Harrie Oosterhuis}.}
  \bibinfo{year}{2022}\natexlab{a}.
\newblock \showarticletitle{Doubly-Robust Estimation for Unbiased
  Learning-to-Rank from Position-Biased Click Feedback}.
\newblock \bibinfo{journal}{\emph{arXiv preprint arXiv:2203.17118}}
  (\bibinfo{year}{2022}).
\newblock


\bibitem[\protect\citeauthoryear{Oosterhuis}{Oosterhuis}{2022b}]%
        {oosterhuis2022reaching}
\bibfield{author}{\bibinfo{person}{Harrie Oosterhuis}.}
  \bibinfo{year}{2022}\natexlab{b}.
\newblock \showarticletitle{Reaching the End of Unbiasedness: Uncovering
  Implicit Limitations of Click-Based Learning to Rank}. In
  \bibinfo{booktitle}{\emph{Proceedings of the 2022 ACM SIGIR International
  Conference on the Theory of Information Retrieval. ACM}}.
\newblock


\bibitem[\protect\citeauthoryear{Oosterhuis}{Oosterhuis}{2023}]%
        {oosterhuis2023doubly}
\bibfield{author}{\bibinfo{person}{Harrie Oosterhuis}.}
  \bibinfo{year}{2023}\natexlab{}.
\newblock \showarticletitle{Doubly Robust Estimation for Correcting Position
  Bias in Click Feedback for Unbiased Learning to Rank}.
\newblock \bibinfo{journal}{\emph{ACM Transactions on Information Systems}}
  \bibinfo{volume}{41}, \bibinfo{number}{3} (\bibinfo{year}{2023}),
  \bibinfo{pages}{1--33}.
\newblock


\bibitem[\protect\citeauthoryear{Oosterhuis and de~Rijke}{Oosterhuis and
  de~Rijke}{2018}]%
        {oosterhuis2018differentiable}
\bibfield{author}{\bibinfo{person}{Harrie Oosterhuis} {and}
  \bibinfo{person}{Maarten de Rijke}.} \bibinfo{year}{2018}\natexlab{}.
\newblock \showarticletitle{Differentiable unbiased online learning to rank}.
  In \bibinfo{booktitle}{\emph{Proceedings of the 27th ACM international
  conference on information and knowledge management}}.
  \bibinfo{pages}{1293--1302}.
\newblock


\bibitem[\protect\citeauthoryear{Oosterhuis and de~Rijke}{Oosterhuis and
  de~Rijke}{2020a}]%
        {oosterhuis2020policy}
\bibfield{author}{\bibinfo{person}{Harrie Oosterhuis} {and}
  \bibinfo{person}{Maarten de Rijke}.} \bibinfo{year}{2020}\natexlab{a}.
\newblock \showarticletitle{Policy-aware Unbiased Learning to Rank for Top-k
  Rankings}. In \bibinfo{booktitle}{\emph{Proceedings of the 43rd International
  ACM SIGIR Conference on Research and Development in Information Retrieval}}.
  \bibinfo{pages}{489--498}.
\newblock


\bibitem[\protect\citeauthoryear{Oosterhuis and de~Rijke}{Oosterhuis and
  de~Rijke}{2020b}]%
        {oosterhuis2020taking}
\bibfield{author}{\bibinfo{person}{Harrie Oosterhuis} {and}
  \bibinfo{person}{Maarten de Rijke}.} \bibinfo{year}{2020}\natexlab{b}.
\newblock \showarticletitle{Taking the Counterfactual Online: Efficient and
  Unbiased Online Evaluation for Ranking}. In
  \bibinfo{booktitle}{\emph{Proceedings of the 2020 ACM SIGIR on International
  Conference on Theory of Information Retrieval}}. \bibinfo{pages}{137--144}.
\newblock


\bibitem[\protect\citeauthoryear{Oosterhuis and de~Rijke}{Oosterhuis and
  de~Rijke}{2021}]%
        {oosterhuis2021unifying}
\bibfield{author}{\bibinfo{person}{Harrie Oosterhuis} {and}
  \bibinfo{person}{Maarten de Rijke}.} \bibinfo{year}{2021}\natexlab{}.
\newblock \showarticletitle{Unifying Online and Counterfactual Learning to
  Rank: A Novel Counterfactual Estimator that Effectively Utilizes Online
  Interventions}. In \bibinfo{booktitle}{\emph{Proceedings of the 14th ACM
  International Conference on Web Search and Data Mining}}.
  \bibinfo{pages}{463--471}.
\newblock


\bibitem[\protect\citeauthoryear{Oosterhuis, Jagerman, and de~Rijke}{Oosterhuis
  et~al\mbox{.}}{2020}]%
        {oosterhuis2020unbiased}
\bibfield{author}{\bibinfo{person}{Harrie Oosterhuis}, \bibinfo{person}{Rolf
  Jagerman}, {and} \bibinfo{person}{Maarten de Rijke}.}
  \bibinfo{year}{2020}\natexlab{}.
\newblock \showarticletitle{Unbiased Learning to Rank: Counterfactual and
  Online Approaches}. In \bibinfo{booktitle}{\emph{Companion Proceedings of the
  Web Conference 2020}}. \bibinfo{pages}{299--300}.
\newblock


\bibitem[\protect\citeauthoryear{Ovaisi, Ahsan, Zhang, Vasilaky, and
  Zheleva}{Ovaisi et~al\mbox{.}}{2020}]%
        {ovaisi2020correcting}
\bibfield{author}{\bibinfo{person}{Zohreh Ovaisi}, \bibinfo{person}{Ragib
  Ahsan}, \bibinfo{person}{Yifan Zhang}, \bibinfo{person}{Kathryn Vasilaky},
  {and} \bibinfo{person}{Elena Zheleva}.} \bibinfo{year}{2020}\natexlab{}.
\newblock \showarticletitle{Correcting for selection bias in learning-to-rank
  systems}. In \bibinfo{booktitle}{\emph{Proceedings of The Web Conference
  2020}}. \bibinfo{pages}{1863--1873}.
\newblock


\bibitem[\protect\citeauthoryear{Ovaisi, Vasilaky, and Zheleva}{Ovaisi
  et~al\mbox{.}}{2021}]%
        {ovaisi2021propensity}
\bibfield{author}{\bibinfo{person}{Zohreh Ovaisi}, \bibinfo{person}{Kathryn
  Vasilaky}, {and} \bibinfo{person}{Elena Zheleva}.}
  \bibinfo{year}{2021}\natexlab{}.
\newblock \showarticletitle{Propensity-Independent Bias Recovery in Offline
  Learning-to-Rank Systems}. In \bibinfo{booktitle}{\emph{Proceedings of the
  44th International ACM SIGIR Conference on Research and Development in
  Information Retrieval}}. \bibinfo{pages}{1763--1767}.
\newblock


\bibitem[\protect\citeauthoryear{Saito}{Saito}{2020}]%
        {saito2020doubly}
\bibfield{author}{\bibinfo{person}{Yuta Saito}.}
  \bibinfo{year}{2020}\natexlab{}.
\newblock \showarticletitle{Doubly robust estimator for ranking metrics with
  post-click conversions}. In \bibinfo{booktitle}{\emph{Proceedings of the 14th
  ACM Conference on Recommender Systems}}. \bibinfo{pages}{92--100}.
\newblock


\bibitem[\protect\citeauthoryear{Saito and Joachims}{Saito and
  Joachims}{2021}]%
        {saito2021counterfactual}
\bibfield{author}{\bibinfo{person}{Yuta Saito} {and} \bibinfo{person}{Thorsten
  Joachims}.} \bibinfo{year}{2021}\natexlab{}.
\newblock \showarticletitle{Counterfactual Learning and Evaluation for
  Recommender Systems: Foundations, Implementations, and Recent Advances}. In
  \bibinfo{booktitle}{\emph{Fifteenth ACM Conference on Recommender Systems}}.
  \bibinfo{pages}{828--830}.
\newblock


\bibitem[\protect\citeauthoryear{Saito and Joachims}{Saito and
  Joachims}{2022a}]%
        {saito2022counterfactual}
\bibfield{author}{\bibinfo{person}{Yuta Saito} {and} \bibinfo{person}{Thorsten
  Joachims}.} \bibinfo{year}{2022}\natexlab{a}.
\newblock \showarticletitle{Counterfactual Evaluation and Learning for
  Interactive Systems: Foundations, Implementations, and Recent Advances}. In
  \bibinfo{booktitle}{\emph{Proceedings of the 28th ACM SIGKDD Conference on
  Knowledge Discovery and Data Mining}}. \bibinfo{pages}{4824--4825}.
\newblock


\bibitem[\protect\citeauthoryear{Saito and Joachims}{Saito and
  Joachims}{2022b}]%
        {saito2022off}
\bibfield{author}{\bibinfo{person}{Yuta Saito} {and} \bibinfo{person}{Thorsten
  Joachims}.} \bibinfo{year}{2022}\natexlab{b}.
\newblock \showarticletitle{Off-policy evaluation for large action spaces via
  embeddings}.
\newblock \bibinfo{journal}{\emph{arXiv preprint arXiv:2202.06317}}
  (\bibinfo{year}{2022}).
\newblock


\bibitem[\protect\citeauthoryear{Sanderson, Paramita, Clough, and
  Kanoulas}{Sanderson et~al\mbox{.}}{2010}]%
        {sanderson2010user}
\bibfield{author}{\bibinfo{person}{Mark Sanderson},
  \bibinfo{person}{Monica~Lestari Paramita}, \bibinfo{person}{Paul Clough},
  {and} \bibinfo{person}{Evangelos Kanoulas}.} \bibinfo{year}{2010}\natexlab{}.
\newblock \showarticletitle{Do User Preferences and Evaluation Measures Line
  Up?}. In \bibinfo{booktitle}{\emph{Proceedings of the 33rd international ACM
  SIGIR conference on Research and development in information retrieval}}.
  \bibinfo{pages}{555--562}.
\newblock


\bibitem[\protect\citeauthoryear{Sarvi, Heuss, Aliannejadi, Schelter, and
  de~Rijke}{Sarvi et~al\mbox{.}}{2021}]%
        {sarvi2021understanding}
\bibfield{author}{\bibinfo{person}{Fatemeh Sarvi}, \bibinfo{person}{Maria
  Heuss}, \bibinfo{person}{Mohammad Aliannejadi}, \bibinfo{person}{Sebastian
  Schelter}, {and} \bibinfo{person}{Maarten de Rijke}.}
  \bibinfo{year}{2021}\natexlab{}.
\newblock \showarticletitle{Understanding and Mitigating the Effect of Outliers
  in Fair Ranking}.
\newblock \bibinfo{journal}{\emph{arXiv preprint arXiv:2112.11251}}
  (\bibinfo{year}{2021}).
\newblock


\bibitem[\protect\citeauthoryear{Schuth, Oosterhuis, Whiteson, and
  de~Rijke}{Schuth et~al\mbox{.}}{2016}]%
        {schuth2016multileave}
\bibfield{author}{\bibinfo{person}{Anne Schuth}, \bibinfo{person}{Harrie
  Oosterhuis}, \bibinfo{person}{Shimon Whiteson}, {and}
  \bibinfo{person}{Maarten de Rijke}.} \bibinfo{year}{2016}\natexlab{}.
\newblock \showarticletitle{Multileave gradient descent for fast online
  learning to rank}. In \bibinfo{booktitle}{\emph{proceedings of the ninth ACM
  international conference on web search and data mining}}.
  \bibinfo{pages}{457--466}.
\newblock


\bibitem[\protect\citeauthoryear{Singh and Joachims}{Singh and
  Joachims}{2018}]%
        {singh2018fairness}
\bibfield{author}{\bibinfo{person}{Ashudeep Singh} {and}
  \bibinfo{person}{Thorsten Joachims}.} \bibinfo{year}{2018}\natexlab{}.
\newblock \showarticletitle{Fairness of exposure in rankings}. In
  \bibinfo{booktitle}{\emph{Proceedings of the 24th ACM SIGKDD International
  Conference on Knowledge Discovery \& Data Mining}}.
  \bibinfo{pages}{2219--2228}.
\newblock


\bibitem[\protect\citeauthoryear{Singh and Joachims}{Singh and
  Joachims}{2019}]%
        {singh2019policy}
\bibfield{author}{\bibinfo{person}{Ashudeep Singh} {and}
  \bibinfo{person}{Thorsten Joachims}.} \bibinfo{year}{2019}\natexlab{}.
\newblock \showarticletitle{Policy Learning for Fairness in Ranking}.
\newblock \bibinfo{journal}{\emph{arXiv preprint arXiv:1902.04056}}
  (\bibinfo{year}{2019}).
\newblock


\bibitem[\protect\citeauthoryear{Singh, Kempe, and Joachims}{Singh
  et~al\mbox{.}}{2021}]%
        {singh2021fairness}
\bibfield{author}{\bibinfo{person}{Ashudeep Singh}, \bibinfo{person}{David
  Kempe}, {and} \bibinfo{person}{Thorsten Joachims}.}
  \bibinfo{year}{2021}\natexlab{}.
\newblock \showarticletitle{Fairness in ranking under uncertainty}.
\newblock \bibinfo{journal}{\emph{Advances in Neural Information Processing
  Systems}}  \bibinfo{volume}{34} (\bibinfo{year}{2021}),
  \bibinfo{pages}{11896--11908}.
\newblock


\bibitem[\protect\citeauthoryear{Tian, Guo, Ostuni, and Zhu}{Tian
  et~al\mbox{.}}{2020}]%
        {tian2020hte}
\bibfield{author}{\bibinfo{person}{Mucun Tian}, \bibinfo{person}{Chu Guo},
  \bibinfo{person}{Vito~Claudio Ostuni}, {and} \bibinfo{person}{Zhen Zhu}.}
  \bibinfo{year}{2020}\natexlab{}.
\newblock \showarticletitle{Counterfactual Learning to Rank using Heterogeneous
  Treatment Effect Estimation}.
\newblock \bibinfo{journal}{\emph{ArXiv}}  \bibinfo{volume}{abs/2007.09798}
  (\bibinfo{year}{2020}).
\newblock


\bibitem[\protect\citeauthoryear{Vardasbi, de~Rijke, and Markov}{Vardasbi
  et~al\mbox{.}}{2020a}]%
        {vardasbi2020cascade}
\bibfield{author}{\bibinfo{person}{Ali Vardasbi}, \bibinfo{person}{Maarten de
  Rijke}, {and} \bibinfo{person}{Ilya Markov}.}
  \bibinfo{year}{2020}\natexlab{a}.
\newblock \showarticletitle{Cascade model-based propensity estimation for
  counterfactual learning to rank}. In \bibinfo{booktitle}{\emph{Proceedings of
  the 43rd International ACM SIGIR Conference on Research and Development in
  Information Retrieval}}. \bibinfo{pages}{2089--2092}.
\newblock


\bibitem[\protect\citeauthoryear{Vardasbi, Oosterhuis, and de~Rijke}{Vardasbi
  et~al\mbox{.}}{2020b}]%
        {vardasbi2020inverse}
\bibfield{author}{\bibinfo{person}{Ali Vardasbi}, \bibinfo{person}{Harrie
  Oosterhuis}, {and} \bibinfo{person}{Maarten de Rijke}.}
  \bibinfo{year}{2020}\natexlab{b}.
\newblock \showarticletitle{When Inverse Propensity Scoring does not Work:
  Affine Corrections for Unbiased Learning to Rank}. In
  \bibinfo{booktitle}{\emph{Proceedings of the 29th ACM International
  Conference on Information \& Knowledge Management}}.
  \bibinfo{pages}{1475--1484}.
\newblock


\bibitem[\protect\citeauthoryear{Vardasbi, Sarvi, and de~Rijke}{Vardasbi
  et~al\mbox{.}}{2022}]%
        {vardasbi2022probabilistic}
\bibfield{author}{\bibinfo{person}{Ali Vardasbi}, \bibinfo{person}{Fatemeh
  Sarvi}, {and} \bibinfo{person}{Maarten de Rijke}.}
  \bibinfo{year}{2022}\natexlab{}.
\newblock \showarticletitle{Probabilistic Permutation Graph Search: Black-Box
  Optimization for Fairness in Ranking}. In
  \bibinfo{booktitle}{\emph{Proceedings of the 45th International ACM SIGIR
  Conference on Research and Development in Information Retrieval}} (Madrid,
  Spain) \emph{(\bibinfo{series}{SIGIR '22})}. \bibinfo{publisher}{Association
  for Computing Machinery}, \bibinfo{address}{New York, NY, USA},
  \bibinfo{pages}{715–725}.
\newblock
\showISBNx{9781450387323}
\urldef\tempurl%
\url{https://doi.org/10.1145/3477495.3532045}
\showDOI{\tempurl}


\bibitem[\protect\citeauthoryear{Wang, Bendersky, Metzler, and Najork}{Wang
  et~al\mbox{.}}{2016}]%
        {wang2016learning}
\bibfield{author}{\bibinfo{person}{Xuanhui Wang}, \bibinfo{person}{Michael
  Bendersky}, \bibinfo{person}{Donald Metzler}, {and} \bibinfo{person}{Marc
  Najork}.} \bibinfo{year}{2016}\natexlab{}.
\newblock \showarticletitle{Learning to Rank with Selection Bias in Personal
  Search}. In \bibinfo{booktitle}{\emph{Proceedings of the 39th International
  ACM SIGIR conference on Research and Development in Information Retrieval}}.
  ACM, \bibinfo{pages}{115--124}.
\newblock


\bibitem[\protect\citeauthoryear{Wang, Golbandi, Bendersky, Metzler, and
  Najork}{Wang et~al\mbox{.}}{2018}]%
        {wang2018position}
\bibfield{author}{\bibinfo{person}{Xuanhui Wang}, \bibinfo{person}{Nadav
  Golbandi}, \bibinfo{person}{Michael Bendersky}, \bibinfo{person}{Donald
  Metzler}, {and} \bibinfo{person}{Marc Najork}.}
  \bibinfo{year}{2018}\natexlab{}.
\newblock \showarticletitle{Position Bias Estimation for Unbiased Learning to
  Rank in Personal Search}. In \bibinfo{booktitle}{\emph{Proceedings of the
  Eleventh ACM International Conference on Web Search and Data Mining}}.
  \bibinfo{pages}{610--618}.
\newblock


\bibitem[\protect\citeauthoryear{Wang, Zhang, Sun, and Qi}{Wang
  et~al\mbox{.}}{2019}]%
        {wang2019doubly}
\bibfield{author}{\bibinfo{person}{Xiaojie Wang}, \bibinfo{person}{Rui Zhang},
  \bibinfo{person}{Yu Sun}, {and} \bibinfo{person}{Jianzhong Qi}.}
  \bibinfo{year}{2019}\natexlab{}.
\newblock \showarticletitle{Doubly robust joint learning for recommendation on
  data missing not at random}. In \bibinfo{booktitle}{\emph{International
  Conference on Machine Learning}}. PMLR, \bibinfo{pages}{6638--6647}.
\newblock


\bibitem[\protect\citeauthoryear{Wu, Chen, Zhao, He, Yin, and Chang}{Wu
  et~al\mbox{.}}{2021}]%
        {wu2021unbiased}
\bibfield{author}{\bibinfo{person}{Xinwei Wu}, \bibinfo{person}{Hechang Chen},
  \bibinfo{person}{Jiashu Zhao}, \bibinfo{person}{Li He},
  \bibinfo{person}{Dawei Yin}, {and} \bibinfo{person}{Yi Chang}.}
  \bibinfo{year}{2021}\natexlab{}.
\newblock \showarticletitle{Unbiased learning to rank in feeds recommendation}.
  In \bibinfo{booktitle}{\emph{Proceedings of the 14th ACM International
  Conference on Web Search and Data Mining}}. \bibinfo{pages}{490--498}.
\newblock


\bibitem[\protect\citeauthoryear{Yadav, Du, and Joachims}{Yadav
  et~al\mbox{.}}{2021}]%
        {yadav2021policy}
\bibfield{author}{\bibinfo{person}{Himank Yadav}, \bibinfo{person}{Zhengxiao
  Du}, {and} \bibinfo{person}{Thorsten Joachims}.}
  \bibinfo{year}{2021}\natexlab{}.
\newblock \showarticletitle{Policy-Gradient Training of Fair and Unbiased
  Ranking Functions}. In \bibinfo{booktitle}{\emph{Proceedings of the 44th
  International ACM SIGIR Conference on Research and Development in Information
  Retrieval}}. \bibinfo{pages}{1044--1053}.
\newblock


\bibitem[\protect\citeauthoryear{Yan, Qin, Zhuang, Wang, Bendersky, and
  Najork}{Yan et~al\mbox{.}}{2022}]%
        {yan2022revisiting}
\bibfield{author}{\bibinfo{person}{Le Yan}, \bibinfo{person}{Zhen Qin},
  \bibinfo{person}{Honglei Zhuang}, \bibinfo{person}{Xuanhui Wang},
  \bibinfo{person}{Mike Bendersky}, {and} \bibinfo{person}{Marc Najork}.}
  \bibinfo{year}{2022}\natexlab{}.
\newblock \showarticletitle{Revisiting two tower models for unbiased learning
  to rank}.
\newblock  (\bibinfo{year}{2022}).
\newblock


\bibitem[\protect\citeauthoryear{Yang and Ai}{Yang and Ai}{2021}]%
        {yang2021maximizing}
\bibfield{author}{\bibinfo{person}{Tao Yang} {and} \bibinfo{person}{Qingyao
  Ai}.} \bibinfo{year}{2021}\natexlab{}.
\newblock \showarticletitle{Maximizing Marginal Fairness for Dynamic Learning
  to Rank}. In \bibinfo{booktitle}{\emph{Proceedings of the Web Conference
  2021}} (Ljubljana, Slovenia) \emph{(\bibinfo{series}{WWW '21})}.
  \bibinfo{publisher}{Association for Computing Machinery},
  \bibinfo{address}{New York, NY, USA}, \bibinfo{pages}{137–145}.
\newblock
\showISBNx{9781450383127}
\urldef\tempurl%
\url{https://doi.org/10.1145/3442381.3449901}
\showDOI{\tempurl}


\bibitem[\protect\citeauthoryear{Yue and Joachims}{Yue and Joachims}{2009}]%
        {yue2009interactively}
\bibfield{author}{\bibinfo{person}{Yisong Yue} {and} \bibinfo{person}{Thorsten
  Joachims}.} \bibinfo{year}{2009}\natexlab{}.
\newblock \showarticletitle{Interactively optimizing information retrieval
  systems as a dueling bandits problem}. In
  \bibinfo{booktitle}{\emph{Proceedings of the 26th Annual International
  Conference on Machine Learning}}. \bibinfo{pages}{1201--1208}.
\newblock


\bibitem[\protect\citeauthoryear{Zhao, Xu, Zhang, Cai, Dong, and Wen}{Zhao
  et~al\mbox{.}}{2022}]%
        {zhao2022unbiased}
\bibfield{author}{\bibinfo{person}{Haiyuan Zhao}, \bibinfo{person}{Jun Xu},
  \bibinfo{person}{Xiao Zhang}, \bibinfo{person}{Guohao Cai},
  \bibinfo{person}{Zhenhua Dong}, {and} \bibinfo{person}{Ji-Rong Wen}.}
  \bibinfo{year}{2022}\natexlab{}.
\newblock \showarticletitle{Unbiased Top-k Learning to Rank with Causal
  Likelihood Decomposition}.
\newblock \bibinfo{journal}{\emph{arXiv preprint arXiv:2204.00815}}
  (\bibinfo{year}{2022}).
\newblock


\bibitem[\protect\citeauthoryear{Zheng, Qiu, Xu, Wu, Zhao, Chen, and
  Xiong}{Zheng et~al\mbox{.}}{2022}]%
        {zheng2022cbr}
\bibfield{author}{\bibinfo{person}{Zhi Zheng}, \bibinfo{person}{Zhaopeng Qiu},
  \bibinfo{person}{Tong Xu}, \bibinfo{person}{Xian Wu},
  \bibinfo{person}{Xiangyu Zhao}, \bibinfo{person}{Enhong Chen}, {and}
  \bibinfo{person}{Hui Xiong}.} \bibinfo{year}{2022}\natexlab{}.
\newblock \showarticletitle{CBR: Context Bias aware Recommendation for
  Debiasing User Modeling and Click Prediction}. In
  \bibinfo{booktitle}{\emph{Proceedings of the ACM Web Conference 2022}}.
  \bibinfo{pages}{2268--2276}.
\newblock


\bibitem[\protect\citeauthoryear{Zhuang, Qin, Wang, Bendersky, Qian, Hu, and
  Chen}{Zhuang et~al\mbox{.}}{2021}]%
        {zhuang2021cross}
\bibfield{author}{\bibinfo{person}{Honglei Zhuang}, \bibinfo{person}{Zhen Qin},
  \bibinfo{person}{Xuanhui Wang}, \bibinfo{person}{Michael Bendersky},
  \bibinfo{person}{Xinyu Qian}, \bibinfo{person}{Po Hu}, {and}
  \bibinfo{person}{Dan~Chary Chen}.} \bibinfo{year}{2021}\natexlab{}.
\newblock \showarticletitle{Cross-positional attention for debiasing clicks}.
  In \bibinfo{booktitle}{\emph{Proceedings of the Web Conference 2021}}.
  \bibinfo{pages}{788--797}.
\newblock


\end{thebibliography}
\end{document}